\documentclass[apj,myRevtex4]{myemulateapj}
\usepackage{graphicx}
\usepackage{amssymb,amsmath}
\usepackage{natbib}
\makeatletter

\newcommand{\Rmnum}[1]{\expandafter\@slowromancap\romannumeral #1@}
\makeatother

\shorttitle{Warm {\it Spitzer} Photometry of Transiting Hot Jupiters HAT-P-4\lowercase{b}, HAT-P-4\lowercase{b} and HAT-P-12\lowercase{b}}
\shortauthors{Todorov et al.}
\begin{document}
\title{Warm {\it Spitzer} Photometry of Three Hot Jupiters: 
HAT-P-3\lowercase{b}, HAT-P-4\lowercase{b} and HAT-P-12\lowercase{b}}
\author{
Kamen O. Todorov\altaffilmark{1}$^,$\altaffilmark{2}$^,$\altaffilmark{3},
Drake Deming\altaffilmark{3},
Heather A. Knutson\altaffilmark{4},
Adam Burrows\altaffilmark{5},
Jonathan J. Fortney\altaffilmark{6},
Nikole K. Lewis\altaffilmark{7}$^,$\altaffilmark{8},
Nicolas B. Cowan\altaffilmark{9}, 
Eric Agol\altaffilmark{10}, 
Jean-Michel Desert\altaffilmark{8}$^,$\altaffilmark{11}, 
Pedro V. Sada\altaffilmark{12},
David Charbonneau\altaffilmark{13}, 
Gregory Laughlin\altaffilmark{6},
Jonathan Langton\altaffilmark{14},
Adam P. Showman\altaffilmark{15}
}
\altaffiltext{1}{Department of Astronomy and Astrophysics, The Pennsylvania
State University, University Park, PA~16802, USA}
\altaffiltext{2}{Center for Exoplanets and Habitable Worlds, The Pennsylvania State University, 
University Park, PA~16802, USA}
\altaffiltext{3}{Department of Astronomy, University of Maryland at College Park, 
College Park, MD~20742, USA}
\altaffiltext{4}{Division of Geological and Planetary Sciences, California Institute 
of Technology, Pasadena, CA~91125, USA}
\altaffiltext{5}{Department of Astrophysical Sciences, Princeton University, Princeton, NJ~08544, USA}
\altaffiltext{6}{Department of Astronomy and Astrophysics, University of California at Santa Cruz, Santa Cruz, CA~95064, USA}
\altaffiltext{7}{Department of Earth, Atmospheric, and Planetary Sciences, 
Massachusetts Institute of Technology, Cambridge, MA~02139, USA}
\altaffiltext{8}{Sagan Fellow}
\altaffiltext{9}{CIERA Fellow, Department of Physics \& Astronomy, Northwestern University, Evanston, IL~60208, USA}
\altaffiltext{10}{Department of Astronomy, University of Washington, Box 351580, Seattle, WA~98195, USA}
\altaffiltext{11}{Astronomy Department, California Institute of Technology, Pasadena, CA~91125, USA}
\altaffiltext{12}{University of Monterrey, Monterrey, Mexico}
\altaffiltext{13}{Harvard-Smithsonian Center for Astrophysics, Cambridge, MA~02138, USA}
\altaffiltext{14}{Department of Physics, Principia College, Elsah, IL~62028, USA}
\altaffiltext{15}{Lunar and Planetary Laboratory, University of Arizona, Tucson, AZ~85721, USA}

\begin{abstract}
We present Warm {\it Spitzer/IRAC} secondary eclipse time series photometry 
of three short-period transiting exoplanets, HAT-P-3b, HAT-P-4b and HAT-P-12b, 
in both the available 3.6 and 4.5\,$\mu$m bands. HAT-P-3b and HAT-P-4b are Jupiter-mass, 
objects orbiting an early K and an early G dwarf stars, respectively. 
For HAT-P-3b we find eclipse depths of $0.112\%^{+0.015\%}_{-0.030\%}$ (3.6\,$\mu$m)
and $0.094\%^{+0.016\%}_{-0.009\%}$ (4.5\,$\mu$m). The HAT-P-4b values are 
$0.142\%^{+0.014\%}_{-0.016\%}$ (3.6\,$\mu$m) and $0.122\%^{+0.012\%}_{-0.014\%}$(4.5\,$\mu$m).
The two planets' photometry is consistent with inefficient heat redistribution 
from their day to night sides (and low albedos), but it is inconclusive about possible temperature 
inversions in their atmospheres. 
HAT-P-12b is a Saturn-mass planet and is one of the coolest 
planets ever observed during secondary eclipse, along with hot Neptune GJ~436b
and hot Saturn WASP-29b. We are able to place 3$\sigma$ upper limits on 
the secondary eclipse depth of HAT-P-12b in both wavelengths: 
$< 0.042\%$ (3.6\,$\mu$m) and $<0.085\%$ (4.5\,$\mu$m). 
We discuss these results in the context of the {\it Spitzer} secondary eclipse 
measurements of GJ~436b and WASP-29b.
It is possible that we do not detect the eclipses of HAT-P-12b 
due to high eccentricity, but find that weak planetary emission in these
wavelengths is a more likely explanation.  
We place 3$\sigma$ upper limits on the $|e\,\cos\omega|$ quantity 
(where $e$ is eccentricity and $\omega$ is the argument of periapsis)
for HAT-P-3b ($< 0.0081$) and HAT-P-4b ($< 0.0042$), based on the secondary eclipse timings.

\end{abstract}


\section{Introduction}
\label{sec:intro}

There are about 177 confirmed exoplanets with orbital periods of less than 10 days
and masses greater than 0.1 Jupiter mass, which are often referred to as
``hot Jupiters''. About 145 of them transit their host stars\footnote{Determined 
from the Exoplanet Orbit Database \citep{wri11} at http://exoplanets.org}. 
Hot Jupiters have high equilibrium temperatures, often over 1000\,K, due to 
their proximity to their parent stars. These planets make ideal targets for 
the {\it Spitzer Space Telescope}, which can directly detect their thermal 
emission via time series photometry in the infra-red during secondary eclipse. 

The drop of total light from the planet-star system as the planet moves 
behind the star was first measured  independently by \citet{cha05} and 
\citet{dem05}. By measuring the eclipse depth in several photometric 
bands, one can construct a very low
resolution emergent spectrum of the observed planet in the infrared \citep{cha08, gri08}. 
Comparison of similar spectra with models has suggested that there are two 
types of hot Jupiter atmospheres based on the presence or absence of
temperature inversions in their upper layers 
\citep[e.g.,][]{knu08,mac09,tod10,bee11,dem11,tod12}.

The mechanisms that create such inversions are not well established, but 
it is generally believed that planets with an inverted atmosphere have
an additional opacity source in their atmospheres' upper layers, 
where the pressure falls below $\sim0.01$\,bar \citep{bur08, for08}.
In the past, this opacity source has been suggested to be gas phase TiO 
\citep{hub03,bur07,bur08,for06a,for08}. However, this molecule may form grains
on the night sides and in cold traps deep in the atmospheres on the day sides  
of some hot Jupiters with inversions such as HD~209458b, where 
the pressure-temperature profiles are expected to cross the condensation 
curve of TiO \citep{spi09}. 
In addition, TrES-3 is sufficiently heated to have gas phase TiO in its atmosphere, yet 
has no temperature inversion \citep{fre10}, while XO-1b is too cool to 
maintain it, but appears to have an inverted atmosphere \citep{mac08}. 
On the other hand, a more recent 3D mixing 
study of hot Jupiter atmospheres \citep{per13} has shown that
TiO may stay aloft in HD~209458b's atmosphere due to strong vertical mixing, 
if it forms grains no larger than several microns. According to this investigation, 
TiO in the form of larger particles should be depleted by the day-night cold trap. 
Thus, the role of TiO in the formation of temperature inversions remains 
controversial. 

There are at least two alternative explanations for the presence or absence of
temperature inversions. 
\citet{zah09} suggested that sulfur compounds may account for 
the stratospheric opacity in some hot Jupiter atmospheres. 
Another hypothesis is that the presence or absence of temperature inversions
is correlated with the magnetic activity and related UV flux from the host 
star \citep{knu10}. According to this idea, the increased UV flux received 
by planets orbiting active stars destroys the molecule responsible for the 
formation of temperature inversions.

We can test these hypotheses by building up a large sample of hot Jupiters 
with secondary eclipse measurements and search for correlations with other 
system parameters, such as stellar activity or the Ti abundance in the stellar 
photosphere. Although the {\it Spitzer Space Telescope} exhausted the last of its 
cryogen in May 2009, it still remains the best facility for carrying out 
these observations. The observatory has continued operating at a higher
temperature of approximately 29\,K, cooled by passive radiation. Only 
two photometric bands are still available, the 3.6 and 4.5\,$\mu$m 
channels of the Infra-Red Array Camera \citep[IRAC,][]{faz04}. 
Fortunately, measurements in these two bands are often sufficient to 
constrain the presence or absence of temperature inversion in the 
upper layers of hot Jupiters \citep{knu10}. 

The energy budget is another aspect of transiting hot Jupiters that can 
be studied via secondary eclipse photometry. These planets are 
expected to tidally evolve within $\sim1$\,Gyr of their formation, 
and their rotation periods should become equal to their orbital 
periods, assuming zero eccentricity \citep{cor10}. The transfer 
of heat from the day side to the night side can have strong 
influence on the emergent wavelength dependent flux of the 
day side of the planet, which is measured directly by estimating
the secondary eclipse depth at a given wavelength. Relating 
observations to atmospheric models can place constraints on 
the heat redistribution efficiency and the Bond albedo of the 
planet \citep{cow11}. The authors of this study suggest that 
planets with very high equilibrium temperatures have a narrow range of  
$\rm T_d/T_0$, where $\rm T_d$ is the effective dayside 
temperature and $\rm T_0$ is the equilibrium temperature at 
the substellar point. This ratio is a measure of the 
redistribution efficiency of the atmosphere, which \citet{cow11}
find to be uniformly low for the hottest planets, while cooler
planets appear to exhibit higher range of heat redistribution 
efficiencies and/or albedos.

In this paper we present {\it Warm Spitzer} time series 
photometry of three short period transiting
exoplanets -- HAT-P-3b \citep{tor07}, HAT-P-4b \citep{kov07} and 
HAT-P-12b \citep{har09}. The physical properties of 
the systems are listed in Table~\ref{tab:prop}. In a previous paper 
\citep{tod12}, we focused on {\it Warm Spitzer} secondary eclipse 
photometry of three planets (XO-4b, HAT-P-6b and HAT-P-8b) with masses comparable to Jupiter's, 
orbiting F dwarfs. The magnetic activity of stars as indicated by the Ca II H\&K line
strengths is uncalibrated for effective temperatures over 6200\,K or under 4200\,K 
\citep{noy1984}. Since the host stars of all three of these targets were above or 
close to the 6200\,K boundary, our results did not reliably test the correlation between magnetic 
activity of the host star and the atmospheric inversions proposed by \citet{knu10}, 
which relies on Ca II H\&K line measurements for assessing the magnetic activity of the stars. 

Therefore, in order to better cover the available 
parameter space, for this analysis we have chosen planets that orbit host stars with 
effective temperatures within the 4200---6200\,K range. Our targets are less massive 
than Jupiter (the least massive, HAT-P-12b, has a mass similar to Saturn's), but have radii 
similar to Jupiter's. In addition to the secondary eclipse depth measurements, we combine
their timings with the most precise ephemerides available (HAT-P-3, HAT-P-4: \citet{sad12}, 
HAT-P-12: this paper) to constrain the orbital eccentricity of the planets. 

For this analysis, we update the code developed by \citet{tod12}, to make use 
of full array mode observations (in the cases of HAT-P-4 and HAT-P-12) and 
variable aperture photometry \citep{lew13}. We estimated the uncertainty of
the eclipse depth and timing with the Markov Chain Monte Carlo and prayer-bead algorithms.
It is in principle possible to use this code to analyze all available
{\it Spitzer} secondary eclipse data sets in a single investigation, 
but we find that every data set has peculiarities that need to be 
addressed on individual basis. We analyze and report eclipse results 
for hot Jupiters in groups of three because this allows for the 
efficiency of analyzing data for multiple exoplanets, while retaining 
the ability to cope with the quirks of individual data sets.

In Section~\ref{sec:obs} we present the {\it Spitzer} observations and our 
photometry procedures. The details of the time series analysis and the 
uncertainty estimation are presented in Section~\ref{sec:ana}. We discuss 
our results in the context of previous studies in Section~\ref{sec:dis}.


\section{Observations and Photometry}
\label{sec:obs}
\subsection{Secondary Eclipse Observations with {\it Spitzer}}
\label{sec:spitzerobs}

We observed the secondary eclipses using the Infra-Red Array
Camera (IRAC) on the {\it Spitzer Space Telescope} in both 
the 3.6\,$\rm{\mu}$m and the 4.5\,$\rm{\mu}$m channels. 
The observations on the HAT-P-3 system in both pass-bands 
utilized the IRAC subarray mode, resulting in 
$32\times32$ pixel (39\arcsec$\times39$\arcsec) images, 
centered on the planet's host star. The HAT-P-4 and HAT-P-12 
data are in full array mode, where each image is taken has the 
full IRAC resolution of $256\times256$ pixels (5.2\arcmin$\times$5.2\arcmin).

The subarray mode images are stacked in FITS data cubes 
that contain 64 exposures taken in a sequence. Our HAT-P-3
observations have effective exposure times per image of
1.92\,s in both pass-bands. The full array data have effective exposure 
times of 4.4\,s (HAT-P-4) and 10.4\,s (HAT-P-12) per image 
respectively in both wavelengths. 
Each secondary eclipse observation of HAT-P-3b lasted for 
7\,h 42\,min resulting in 13,760 images (215 data cubes). 
The observations of HAT-P-4b covered 7\,h 38\,min, resulting 
in 3,871 images in each wavelength, while those of HAT-P-12b 
lasted almost as long -- 7\,h 37\,min -- but resulted in 
only 2,097 images per pass-band, due to the longer exposure time. 

Complete information about the time span of the observations 
is presented in Table~\ref{tab:dat}.

\subsection{Photometry and Time Information Extraction}
\label{sec:phot}

Our time-stamp extraction routine is very similar to 
the one used by \citet{tod12}. In all data sets, we perform the 
extraction on the Basic Calibrated Data (BCD) files produced by 
version S18.18.0 of the {\it Spitzer} pipeline. For timing of the photometric 
points, we rely on the $\rm{MJD\_OBS}$ keyword in the FITS headers, 
corrected to indicate the mid-exposure time. 
We convert this time stamp, given in modified Julian date, based on the 
Coordinated Universal Time (UTC) standard ($\rm MJD_{UTC}$), to 
Barycentric Julian Date based on the Terrestrial Time standard 
(BJD$_{\rm TT}$) using Jason Eastman's IDL routine 
{\it get\_spitzer\_bjd} \citep{eas10}. We prefer to use the TT 
standard rather than UTC, because the latter is discontinuous
and has leap seconds introduced occasionally. At the time of 
our observations, $\rm TT \approx UTC + $66.184$\,s$.

We convert the pixel intensities from $\rm MJy/sr$ to electron counts using the
the information provided in the image headers, in order to be able to estimate 
the Poisson noise of the photometry. The data are filtered for energetic particle hits by 
following each pixel through time. This is done in two passes, first flagging 
all pixels 8\,$\sigma$ or more away from a boxcar median through time with width 5.
Their values are replaced with the local boxcar median value.
In the second pass, all values more than 4\,$\sigma$ away from the boxcar 
median through time (again with width 5) are flagged and their values are replaced 
with the local median value. For the HAT-P-3 data, this procedure is performed separately 
for each data cube instead of for the whole time series in order to avoid pixel 
rejection due to the sharp changes in background that occur between the last frame 
from one cube and the first frame from the next. This effect is similar to the one 
seen by \citet{dem11}. The fraction of corrected pixels is about 
0.53\% (HAT-P-3, at 3.6\,$\rm{\mu}$m), 0.12\% (HAT-P-3, at 4.5\,$\rm{\mu}$m), 
0.13\% (HAT-P-4, at both channels), 0.22\% (HAT-P-12, at 3.6\,$\rm{\mu}$m) and 
0.24\% (HAT-P-12, at 4.5\,$\rm{\mu}$m).

We estimate the background flux from the full-array images by creating
a histogram of all pixel values for each frame and fitting a Gaussian function. 
We correct for contamination by field stars and energetic particle hits 
by fitting only to the central portion of the histogram and excluding the regions 
that account for the high photon count pixels. We test for variable background 
across the full-array images by measuring the background based on 30$\times$30, 
50$\times$50, and 100$\times$100 pixel boxes centered on the star-planet system 
(which was always positioned in the center of the array). Removing the background 
from the photometry in this manner produces at most 0.7\% change in the photometric 
scatter around a running median of width 20 of the photometry compared to data 
reduction using the whole array to estimate background, in all data sets. 
This is marginal, therefore, we elect to maximize our background determination precision 
and use the whole array to determine the background levels. 
Since the subarray mode observations of HAT-P-3 result in only $32\times32$ pixel images, 
we exclude a $5\times5$ pixel square centered on the star from the histograms for 
these data, in order to avoid biasing the background estimation to higher values. 

In order to locate the centroid of the stellar point response function (PRF)
we experiment with fitting a two-dimensional Gaussian function to the core of the stellar image
\citep{agol10} and with flux-weighted centroiding \citep[e.g.,][]{cha08,knu08}. 
We find that in most cases there is an improvement in the standard deviation
of the residuals that remain after we subtract our best fit model from the
photometry, if we adopt flux-weighted centroiding. The only exception is 
the HAT-P-3b 4.5$\mu$m data set, where the difference in the resulting scatter is marginal ($\sim0.1\%$).
Therefore, we adopt flux-weighted centroiding for all data sets in this analysis. 

We experiment with two different photometry approaches. 
First, we perform aperture photometry, using the IDL routine 
{\it aper}\footnote{http://idlastro.gsfc.nasa.gov/}, 
varying the aperture radius in increments of 0.5 pixels between 1.5 and 6.5 pixels. 
We select the best photometry aperture radius by measuring the true scatter it produces
around a boxcar median of the raw light curve with width 20. The scatter is not
strongly correlated with aperture radius. For the HAT-P-3 subarray mode data, 
the minimum scatter is found at 2.5px (3.6\,$\mu$m) and 3.0px (4.5\,$\mu$m) pixel radii.
For the full array mode data we find that the smallest scatter occurs at
pixel radii of 4.0px (HAT-P-4, 3.6\,$\mu$m), 3.0px (HAT-P-4, 4.5\,$\mu$m), 
5.0px (HAT-P-12, 3.6\,$\mu$m) and 2.5px (HAT-P-12, 4.5\,$\mu$m).

Separately, we perform photometry on all data by using time-variable 
aperture. For each image, we estimate the noise-pixel parameter
\citep{mig05,knu12,lew13}, which is a measure of the width of the stellar 
point spread function (PSF). It is defined in Section 2.2.2 of the {\it Spitzer/IRAC} 
instrument handbook as:
\begin{equation}
  \tilde{\beta} = \frac{(\Sigma_i{I_i})^2}{\Sigma_i{I_i^2}},
\label{eqn:beta}
\end{equation}
where $I_i$ is the intensity detected by the $i^{th}$ pixel. The noise pixel 
parameter, $\tilde{\beta}$, is proportional to the full-width-half-maximum of 
the stellar PSF \citep{mig05}. For each image, we calculate 
the photometric aperture radius:
\begin{equation}
  r = \sqrt{\tilde{\beta}}b + c,
\end{equation}
where b is a scaling factor and c is a constant. In each image frame, 
we measure the flux used to determine $\tilde{\beta}$ using circular aperture
radii between 1.0 and 6.5 pixels. If any part of a pixel falls within the
aperture radius, it is fully included in the $\tilde{\beta}$ calculation.
For each of these values, we vary 
$b$ and $c$ in steps of 0.05. We fit each of the resulting light curves
with our ``systematics-and-eclipse'' model
and find the combination of photometric parameters that yields the 
smallest standard deviation of the residuals. 

For the 3.6$\mu$m data sets, 
the variable aperture photometry approach yields lower residual scatter values 
(between 3 and 7\%), 
reduces the amplitude of the periodic flux oscillation in the raw data
by about 50\% and reduces the levels of correlated noise in the
residuals after the best fits to the data are subtracted from the photometry. 
This is particularly evident in the HAT-P-12b 3.6$\mu$m light curve, where
fixed aperture photometry yields significantly different eclipse depths 
(between $-0.07\%$ and $0.07\%$, which is above our final
3$\sigma$ eclipse depth limit) based on the radius 
of the photometric aperture and the method for decorrelation of the intra-pixel effect.

\begin{figure*} 
\epsscale{0.8}
\plotone{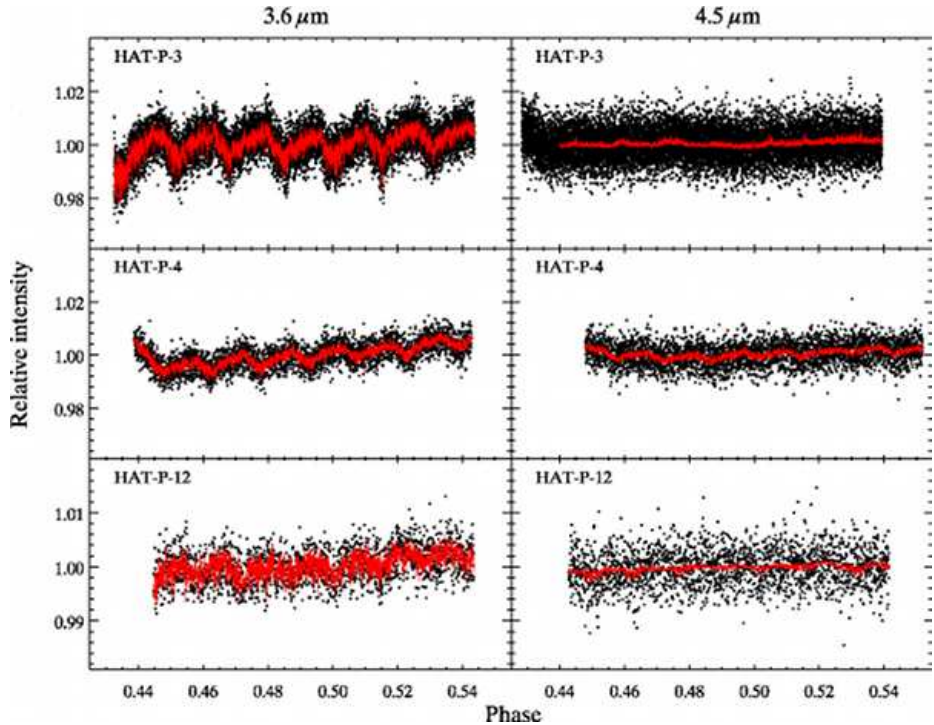}
\caption{
The uncorrected time series photometry for HAT-P-3, HAT-P-4 and HAT-P-12 at 
3.6 and 4.5\,$\mu$m during secondary eclipse (dots), with the best fit eclipse
model we obtain (red lines). These models include the astrophysical eclipse, and
the instrumental effects: a linear ramp and the dependence 
of the measured intensity on the x and y-position of the stellar image on the 
detector. All photometric points are shown here, including the initial 49min of
HAT-P-3 photometry, which we later reject. The best fit models only cover 
data that we have adopted for the fits. 
We discuss the data models in detail in Section~\ref{sec:meas}. The HAT-P-12
photometry is shown on a smaller scale, since these data have the longest 
exposure time, and therefore the highest signal to noise per point. We do not 
detect the secondary eclipses of HAT-P-12b, and in these panels their depths 
are set to zero. 
}
\label{fig:raw}
\end{figure*}

In the 4.5$\mu$m light curves, the residual scatter values are consistently, but marginally 
smaller for the variable aperture than for fixed aperture (between 0.1 and 1\%). There is also 
a slight improvement in the residual correlated noise levels in the light curves   
after decorrelation. Therefore, we adopt variable aperture in the final analysis 
for all data sets. The photometry parameters we used are summarized in 
Table~\ref{tab:phot}. The raw photometry during secondary eclipse for 
the three planets is presented in Figure~\ref{fig:raw}.

\section{Data Analysis}
\label{sec:ana}
\subsection{Ephemerides}
\label{sec:eph}

\begin{figure*} 
\epsscale{0.8}
\plotone{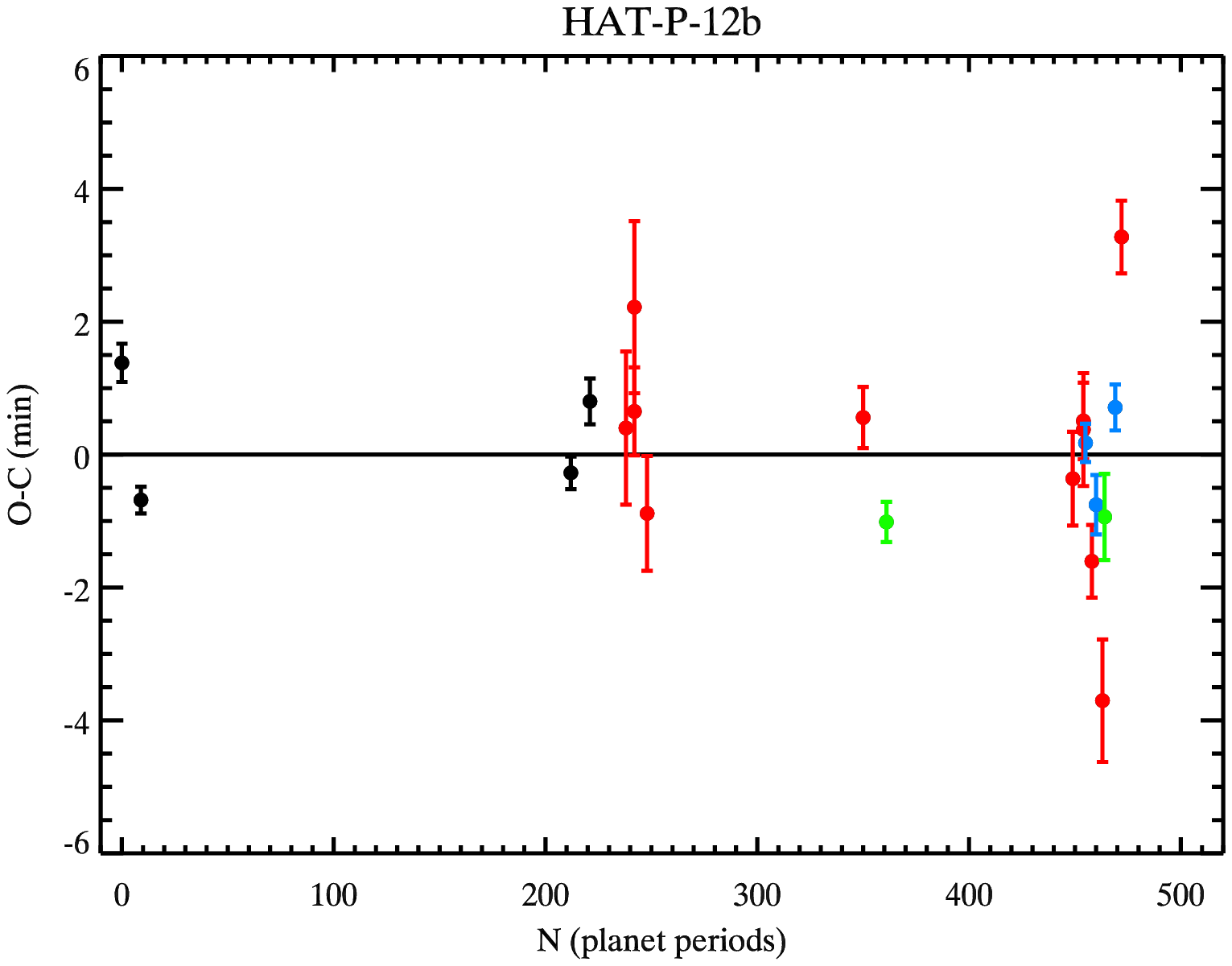}
\caption{
The difference between the best fit transit times and the observed transit times for 
HAT-P-12b. The black symbols represent transit timings from the discovery paper for 
this planet \citep{har09}. Also included are timings from the 
Exoplanet Transit Database (ETD, http://var2.astro.cz/ETD/)
compiled by \citet{lee12} (red symbols), and the measurements made by \citet{lee12}
(blue) and \citet{sad12} (green). The horizontal axis shows the 
number of periods after the $\rm T_0$ transit. We do not see any obvious correlation 
in the residuals that may suggest a transit timing variation, and hence the presence of an 
additional companion in the system. On the other hand, the number of outliers on this plot 
may imply that the timing uncertainties in some of these studies are underestimated. 
}
\label{fig:eph}
\end{figure*}

The data analysis routines fit the eclipse models to the light curves as a function 
of orbital phase. Therefore, the best available estimates of the ephemerides of 
the planets are needed in order to calculate the orbital phase with minimum 
uncertainty. For HAT-P-3b and HAT-P-4b, we use the ephemerides given in 
\citet{sad12}. The ephemeris provided in that paper for HAT-P-12b, however, 
does not include the transits observed recently by \citet{lee12}, and vice versa. 
We combine the transit timings provided by both groups with other transit 
timings from the literature (Table~\ref{tab:eph}). 
The results are $\rm T_0 = 2454187.85559 \pm 0.00011$ in 
$\rm BJD_{TT}$ and $\rm P = 3.21305929 \pm 0.00000034$\,days. The offsets in 
minutes between the best fit ephemeris and the transit timings are shown in 
Figure~\ref{fig:eph}, and they are consistent with an unperturbed orbit. 

\subsection{Secondary Eclipse Fits}
\label{sec:meas}
\subsubsection{Data Examination}

The secondary eclipse depth is only measurable in {\it Spitzer} time series
photometry after careful removal of any instrumental effects. We normalize the
light curve so that the mean brightness of the target system during eclipse 
is unity, corresponding to the light of the star only. We remove any data
that have high backgrounds or are outliers
(9 frames for HAT-P-3 at 3.6\,$\mu$m, 5 frames for HAT-P-3 at 4.5\,$\mu$m, 
22 frames for HAT-P-4 at 3.6\,$\mu$m, 42 frames for HAT-P-4 at 4.5\,$\mu$m, 
75 frames for HAT-P-12 at 3.6\,$\mu$m and 26 frames for HAT-P-12 at 4.5\,$\mu$m).
Like previous investigators \citep[e.g.][]{har07,agol10,dem11,cow12,tod12}, 
we find that the 57$^{\rm th}$ frame in each data cube in the subarray mode exhibits
a relatively high background value, and we exclude all these images from the analysis
(215 frames in each HAT-P-3b band). 

\begin{figure*} 
\epsscale{0.8}
\plotone{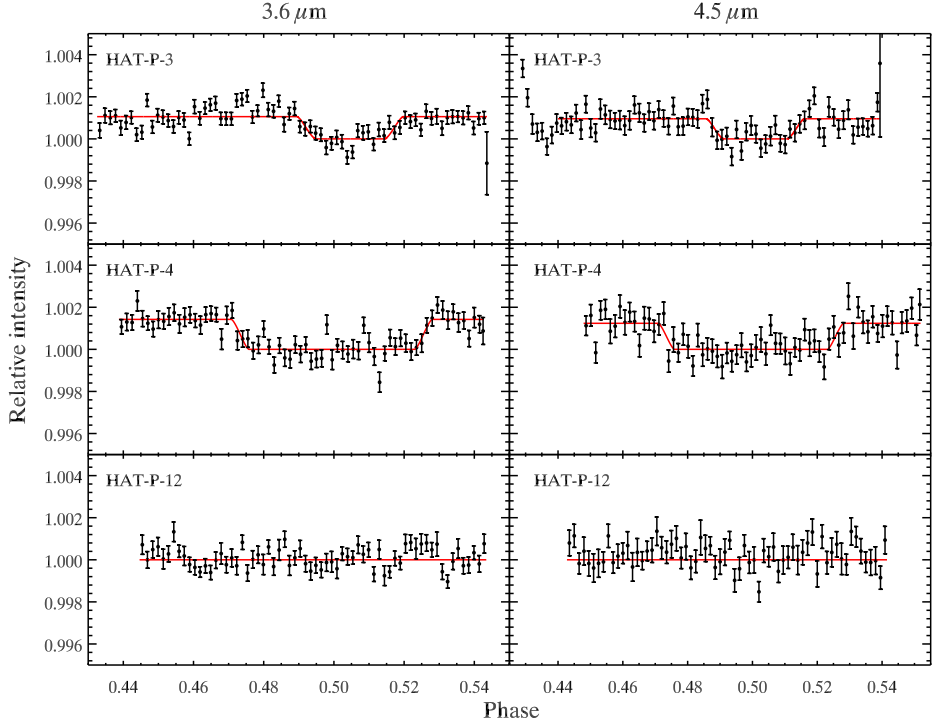}
\caption{
The binned time series photometry for HAT-P-3, HAT-P-4 and HAT-P-12 at 
3.6 and 4.5\,$\mu$m during secondary eclipse after correction for instrumental 
effects (dots). The bin width is 0.0015 in units of orbital phase, corresponding to about 
6\,min 16\,sec (HAP-P-3b), 6\,min 36\,sec (HAT-P-4b) and 6\,min 56\,sec (HAT-P-12b).
The red lines represent the best fit eclipse model. The best fit coefficients
were used to correct all data for instrumental effects, but we have only fitted
to the data covered by the eclipse model. The HAT-P-12b eclipse depths have been 
set to zero in this plot. 
}
\label{fig:bin}
\end{figure*}

Previous time series photometry with the {\it Spitzer/IRAC} instrument, both during the
cryogenic and warm missions, has revealed two different transient instrumental effects. 
First, it often takes tens of minutes for the target star's position on the
detector to stabilize \citep[e.g.,][]{and11}. This initial position 
instability causes apparent changes in intensity because it involves 
portions of the detector that have different intrapixel sensitivity variations 
than the ones used the most during the observation. The second effect, seen 
by e.g., \citet{cam11,dem11,tod12}, causes the apparent brightness at the
start of some observations at 3.6\,$\mu$m to increase or decrease in an 
exponential-like manner before stabilizing, without correlation with the 
position of the stellar image on the detector. This behavior is similar 
to that observed in the longer-wavelength IRAC arrays and is believed to be 
due to charge-trapping \citep[e.g., ][]{knu07, agol10}. The simplest way of 
correcting for these effects is to clip the initial portion of the time series. 
Therefore, we discard the initial 48\,min 42\,sec from the HAT-P-3 data at 
4.5\,$\mu$m, corresponding to 1423 frames taken before orbital phase of 0.44. 
We do not find it necessary to clip the HAT-P-3 at 3.6\,$\mu$m data, or the
HAT-P-4 and HAT-P-12 time series in either wavelength. The corrected light curves 
are shown in Figure~\ref{fig:bin}.


\subsubsection{Initial Fitting Procedure}
\label{sec:fp}

Our procedure for determining the most suitable systematics and eclipse model is similar to that of 
\citet{tod12} -- we assume a central phase of the eclipse and perform a 
simultaneous linear regression fit for all free parameters to the unbinned light curve.
We make incremental increases to the initial assumed phase to scan a wide range of 
possible central phases of the eclipse, and fit a new model to the data, after each
step in phase. We track the $\chi^2$ values of the best regression fits as we make 
the scan. The step size in phase is $10^{-5}$ for all data sets, and we cover 
the phase intervals between 0.48 and 0.52. We experiment with larger phase ranges, especially
in the case of HAT-P-12b, but without an improvement in the results. 

We employ a computational model describing the systematic and astrophysical 
effects observed in our data sets similar to the one used by \citet{tod12}. 
Similarly to previous studies \citep[e.g., ][]{knu09, bee11, dem11}, we find a correlation between 
the X and Y position of the stellar image on the pixels and the measured brightness of 
the star. As in \citet{tod12}, for a given eclipse central phase, we adopt a 
model of the systematics with a quadratic dependence between X and Y in 
intrapixel coordinates without cross-terms and the intensity and a linear ramp with time: 
\begin{equation}
 I(t) = at + b_1X + b_2X^2 + c_1Y + c_2Y^2 + I_0 + e_1M,  
\label{eqn:ipf}
\end{equation} 
where I(t) is the intensity as a function of time, t is time in units of phase, X and Y are
the positions of the stellar centroid on the pixel in the x and y directions, 
M eclipse shape model, and $a, b_1, b_2, c_1, c_2, I_{0}$ (the ordinate axis intercept) and 
$e_{1}$ (the eclipse depth) are the free parameters. We also experimented with adding
the noise pixel parameter, $\tilde{\beta}$, as a third dimension in the spatial fit, but we noticed no 
improvement in the quality of the fits. Thus, we have not included $\tilde{\beta}$ in the fit. 
Multiple previous studies have also settled on quadratic X and Y decorrelation 
\citep[e.g., ][]{cha08,knu08,chr10,and11,coc11,dmr11,des11}. We experiment by setting
all combinations of $a, b_2$ and $c_2,$ to 0 and examine the residuals, but we find
that the lowest residual scatter occurs when all parameters are left free (as expected).
 
We, therefore, attempt to use the Bayesian information criterion test (BIC) to determine the optimal
number of free parameters. We find that for HAT-P-3b (both wavelengths) and HAT-P-12b 
at 3.6\,$\mu$m setting $b_2=0$ results in minimum BIC. The BIC for HAT-P-4b at 4.5\,$\mu$m 
is minimized by setting all parameters free. The minimum BICs for HAT-P-4b at 3.6\,$\mu$m and HAT-P-12b 
at 4.5\,$\mu$m are found when $b_2$ and $c_2,$ are set to 0. However, for these data sets, this results in 
red noise with large amplitude in the residuals, and an unrealistically large eclipse 
depth value for HAT-P-4b at 3.6\,$\mu$m. Setting $b_2$ and $c_2,$ free minimizes the
red noise amplitudes in these light curves. For all other data sets,
regardless of the choice of parameters, the eclipse depth and central phase values are 
within one sigma of each other. We experiment with adding higher order terms to Equation~\ref{eqn:ipf}, but this 
does not lead to improvements in the red noise reduction or the BIC values, and to only to 
marginal improvements in the scatter of the residuals. Therefore, we conclude that the 
BIC test is not ideally suited for the correction of the systematic noise in our data. 

To further motivate this, suppose we found two curves that accounted for the intra-pixel effect, 
and their best-fits were essentially the same curve (i.e. they lay on top of each other), 
but one used many more parameters.  We certainly would not be justified in identifying 
the additional parameters with the physical properties of the detector (lacking a priori 
information on the physics of the effect).  However, as far as removing the effect from 
the photometry, we could use either curve because they would produce 
the same decorrelated photometry, although their BIC values might be very different.  
Thus, the BIC is not necessarily relevant, because we're not seeking information about the detector.
The BIC would be valuable in distinguishing various decorrelation models if they did not
overlap each other, and one curve produced a significantly smaller $\chi^2$ value. However, in our 
data adding cubic and higher terms to the decorrelation polynomial produces only marginal 
decreases in the $\chi^2$.

Thus, we elect not to use the BIC to determine the optimal number of free parameters 
and choose to keep all parameters in Equation~\ref{eqn:ipf} free, while not adding any 
higher order terms. 

The measured eclipse depths for the HAT-P-12 data sets are consistent with 
zero. We experiment by removing various portions of the data at the start and the end 
of the observations. The resulting fits have central phases covering the whole 
explored range, and the eclipse depths take small negative or positive values.
We conclude that our photometric precision is insufficient to detect the eclipse in
these data sets and we place upper limits on its depth. 

For completeness, we experiment with substituting the quadratic dependencies of 
intensity on X and Y with a weighting function, as described by \citet{bal10}, 
which essentially multiplies each photometric point by a ``weight'', dependent
on the X and Y positions of the stellar image on the detector. The apparent
brightness of the star is smoothed as a function of X and Y on the IRAC array, 
with smoothing widths $\rm \sigma_x$ and $\rm \sigma_y$. The in-eclipse data
are excluded to prevent the decorrelation from removing it as a systematic effect. 
We optimize the smoothing widths to minimize the scatter in the residuals of the data after subtracting the 
best fit model. For the 3.6\,$\mu$m data we settle on 
$\rm \sigma_x = 0.00425, 0.00638$, and $0.00723$ and 
$\rm \sigma_y = 0.00723, 0.00638$, and $0.01064$ for HAT-P-3b, 
HAT-P-4b and HAT-P-12b, respectively. For the 4.5\,$\mu$m data, 
we use  $\rm \sigma_x = 0.00894, 0.03404$, and $0.03404$ and 
        $\rm \sigma_y = 0.00936, 0.02085$, and $0.03574$, again for HAT-P-3b, 
HAT-P-4b and HAT-P-12b, respectively. In this case, the BIC is not 
an adequate measure to compare the resulting fits, since this is a non-parametric 
correction of systematic effects 
\citep[for more details on the applicability of the BIC, see e.g., ][]{ste12}. 
Therefore, we compare the mean standard deviations of the residuals after subtracting the
models from the photometry. We find that the scatters produced by weighting 
function fits are comparable or slightly larger than the ones from our polynomial fits, 
and using the weighting function decorrelation does not improve the correlated noise
removal. Therefore, we elect to use quadratic fits in X and Y for all data. 


\subsubsection{Best Parameter Values and Uncertainty Estimates}

In order to determine the best parameter values and their uncertainties, 
we utilize two approaches -- Markov Chain Monte Carlo (MCMC) and prayer-bead
Monte Carlo (PBMC). We first discuss our implementation of these algorithms, and 
then focus on our best fit parameter determination approach. 

We implement a Markov Chain Monte Carlo (MCMC) code in order to quantify the
uncertainties on the eclipse depths and central phases that we measure. 
We follow the recipe suggested by \citet{ford05, ford06}, taking steps 
in one parameter at a time, and drawing the steps from a Gaussian probability
distribution. We perform $10^6$ iterations per free
parameter. Here, we add central phase as a formal free
parameter. Before running the main chain, we run several shorter
chains to optimize the most likely step size for each parameter 
result in acceptance rates between 35\% and 55\%. A typical acceptance 
rate for all parameters and all data sets is $\sim45$\%, which is
near the ideal rate suggested by \citet[][and references therein]{ford06}. We run 
the MCMCs for $7\times10^6$ steps. 

\begin{figure*} 
\epsscale{0.8}
\plotone{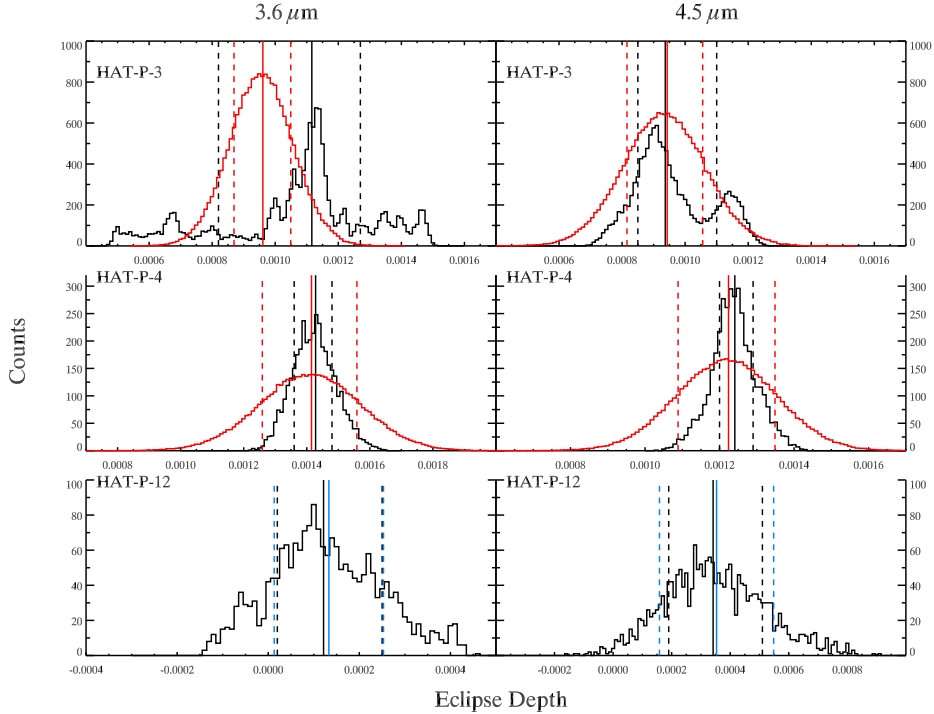}
\caption{
We show the histograms of eclipse depth values that result from the prayer-bead 
technique (black) and the MCMC iterations (red). The vertical solid lines indicate
the best values from the PBMC (black), MCMC (red) and linear regression (blue), 
which we take to be the median of the eclipse depth values of 
the prayer-bead and MCMC runs. The dashed lines
bracket the regions centered on the medians that contain 68\% of the recorded
values for prayer-bead (black) and MCMC (red). Since MCMC and PBMC address 
different noise effects, we adopt as final best values and uncertainties
the results from the technique which yields the larger uncertainty region.
For HAT-P-12, since there is no secondary eclipse detected, 
the MCMC does not converge, and the prayer-bead 
is performed with the central phase fixed at 0.50007. The blue dashed lines 
indicate the $1\sigma$ estimates, based on the linear regression used to determine 
the eclipse depth in the original data sets. We conservatively adopt these as 
final best values since they are larger than the prayer-bead uncertainties.
}
\label{fig:edhist}
\end{figure*}

Using the steps from the Markov chains, we create histograms of the 
eclipse depth and central phase values during a given run. These  
have shapes close to Gaussian (for the eclipse depth histograms, see
Figure~\ref{fig:edhist}) and we estimate the uncertainties on the 
astrophysical parameters by calculating their standard deviations 
from the best fit values. MCMC histograms tend to be close to 
Gaussians even for data sets dominated by correlated noise, since this 
algorithm assumes white noise.

The MCMC fails to converge on a solution for the two HAT-P-12 data sets, 
due to the undetectable eclipses. 
Hence, we report the upper limit for the eclipse depths based on the
regression analysis uncertainty, assuming central phase of 0.50007. This
number includes the light travel time delay assuming $e cos(\omega) = 0$, 
but does not include any apparent delay that may be due to the hottest 
point on the planet trailing behind the substellar point along the orbit.
This effect causes a typically small delay, about $20-30$\,s \citep[e.g., ][]{knu07, agol10}. 

Similarly to, e.g., \citet{des11, dem11, tod12}, 
we obtain an estimate of the systematic uncertainties due to correlated
red noise by performing a ``prayer-bead'' analysis \citep{gil07a}. In this method the 
residuals of the best regression fit are shifted right by one frame 
(last frame becomes first) and added back to the best fit model, thus creating
a simulated data set with the red noise preserved. We perform a 
fit to the simulated data set using the algorithm described in 
Section~\ref{sec:fp}. We record
the resulting eclipse depth and central phase, and simulate another
data set by shifting the residuals of the original best fit again. 
The histograms of the resulting eclipse depth distributions are
presented in Figure~\ref{fig:edhist}. They are non-Gaussian, as 
expected for red noise dominated data sets, but begin to approach 
a Gaussian for data where the white noise dominates. 

The smallest $\chi^2$ value of a fit represents the best fit value 
only in data sets dominated by Gaussian noise. The {\it Spitzer} light curves, 
however, are often dominated by red noise, which is not correlated in time with
the astrophysical signal. Therefore, for a red noise dominated light curve, 
any of the data sets simulated in the PBMC simulation run {\it could} have been the observed data set. 
Hence, for both eclipse depth and central phase, 
we elect to report our best fit value to be the median in the histogram of 
a parameter from the MCMC or PBMC fit simulations, whichever 
yields the larger uncertainty range. We adopt the corresponding uncertainties (Figure~\ref{fig:edhist}). 
If the data sets are dominated by Gaussian noise, our approach
reduces to adopting the parameter values that yield the smallest $\chi^2$.

The best fit eclipse depths from the original data are very close to the medians
from the Monte Carlo runs. For the HAT-P-3 data, the original data best fits
are 0.108\% and 0.096\% versus Monte Carlo median eclipses of $0.112^{+0.015}_{-0.030}$ and $0.094^{+0.016}_{-0.009}$
at 3.6 and 4.5\,$\mu$m, respectively. In both cases, this is a difference of about 0.13$\sigma$.
The HAT-P-4b original data eclipses are 0.142\% and 0.124\% versus $0.142^{+0.014}_{-0.016}$ and
$0.122^{+0.012}_{-0.014}$ from the Monte Carlo runs at 3.6 and 4.5\,$\mu$m, respectively. The
difference between these values is 0$\sigma$ (3.6\,$\mu$m) and 0.17$\sigma$ (4.5\,$\mu$m).
Thus, the exact choice of ``best values'' has no impact on our results. 

We summarize the Monte Carlo results we adopt as final in Table~\ref{tab:ed}. The uncertainties in central 
phase in that table include an uncertainty contribution from the ephemeris, 
but this is small ($\sim1\%$ of the total timing uncertainty). The eclipse
central times in $\rm BJD_{TT}$ are independent of any uncertainty in the 
ephemeris. 

\section{Discussion}
\label{sec:dis}
\subsection{Comparison to Models}
\label{sec:mods}

\begin{figure*} 
\epsscale{0.8}
\plotone{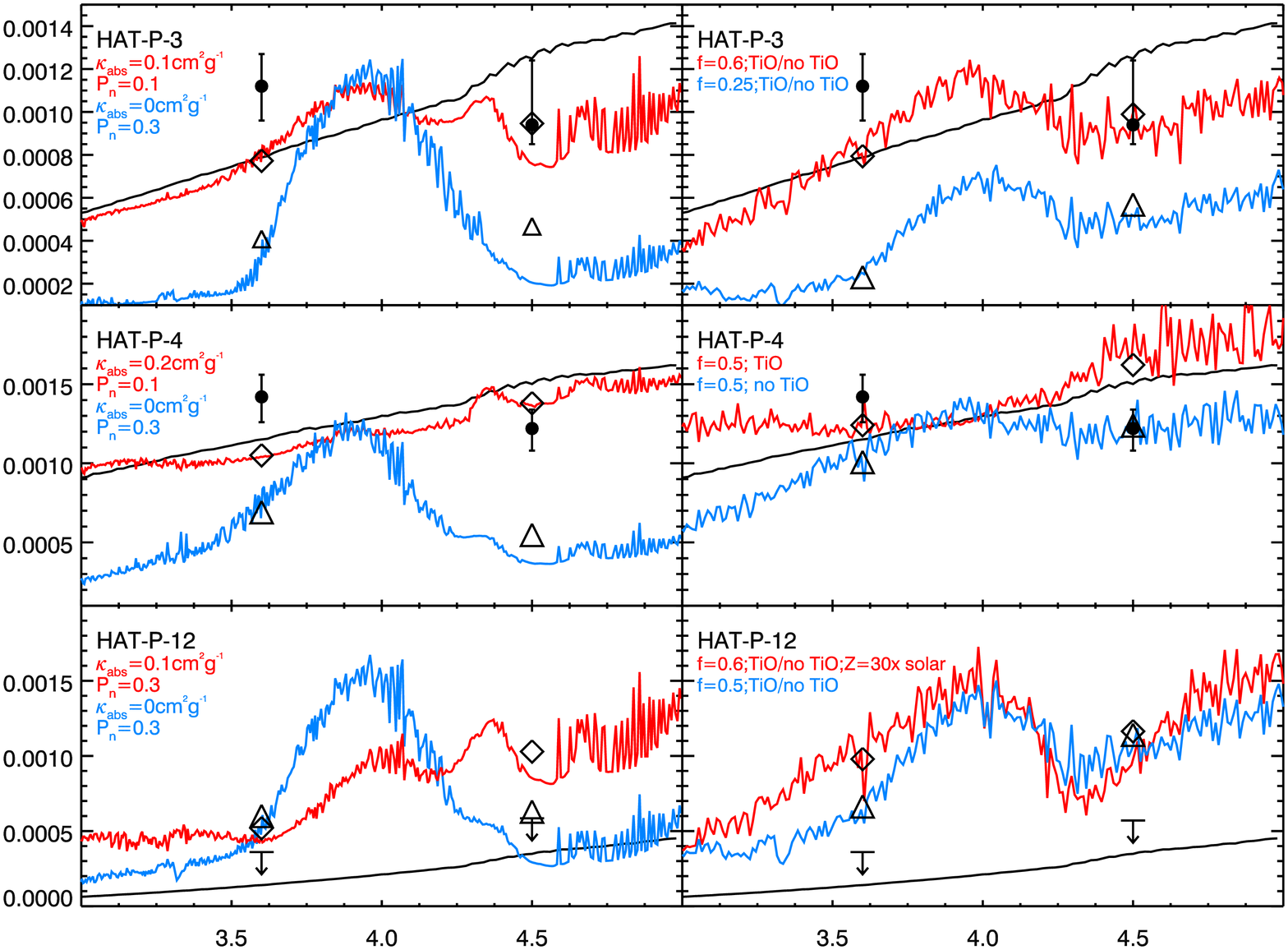}
\caption{
The measured eclipse amplitudes as a function of wavelengths (filled circles) 
compared to different atmospheric models: a blackbody
planet with a Kurucz model for the spectrum of the star \citep[black lines][]{kur79}, 
Burrows planetary atmospheric models \citep[left panels, ][]{bur07,bur08}, and 
similar atmospheric models by \citet[right panels, ][]{for05, for06a,for06b, for08}. 
The downward arrows represent $3\sigma$ upper limits for the HAT-P-12b eclipse depths. 
We have assumed effective stellar temperatures for HAT-P-3 from \citet{chan11}
and from \citet{knu10} for HAT-P-4 and HAT-P-12. Left panels: 
The red lines represent the inverted models, with absorption 
coefficient of the unknown absorber in the upper atmosphere
$\rm \kappa_{abs} = 0.1,~0.2,~$and$~0.1$\,$\rm cm^2\,g^{-1}$ 
and heat redistribution parameter $\rm P_n = 0.1,~0.1,~$and$~0.3$ for
HAT-P-3b, HAT-P-4b and HAT-P-12b, respectively. 
The blue lines are models with no inversion 
($\rm \kappa_{abs} = 0$\,$\rm cm^2\,g^{-1}$) and
$\rm P_n = 0.3$ for all three planets. 
We over-plot the theoretical eclipse depths 
resulting from integrating the model stellar 
and planetary fluxes over the IRAC pass-bands 
(diamonds for the inverted models and triangles 
for the inverted ones).
Right panels: In the Fortney model paradigm, where the stratospheric 
absorbers are assumed to be TiO and VO, HAT-P-3b and HAT-P-12b 
are sufficiently cool, so that the inverted and non-inverted 
(i.e., containing or not TiO and VO in the upper layers of the 
atmosphere) models are indistinguishable, due to TiO and VO condensation and rain-out. 
For HAT-P-3b, we show models with $f = 0.6$ (red line) and 
$f = 0.25$ (blue line); for HAT-P-4b, both models shown have 
$f = 0.5$, but the red model has TiO in the upper layers
of the atmosphere, while the blue one does not. For HAT-P-12b, 
the red line represents a model with $f = 0.6$, while the blue
line shows a model with $f = 0.5$. All Fortney model atmospheres
have solar composition, except the red model for HAT-P-12, which 
has metallicity 30 times higher than solar. For a more detailed 
explanation on the models see Section~\ref{sec:mods}.
}
\label{fig:models}
\end{figure*}

We compare our eclipse depth measurements to two sets of models, 
by \citet{bur07,bur08} and \citet{for05,for06a,for06b, for08},
in order to better understand their implications for the thermal structures
and heat transport efficiency of the planetary atmospheres 
(Figure~\ref{fig:models}). The chemical equilibrium and opacities in the Burrows models
are based on studies by \citet{bur99}
and \citet{sha07}, respectively. The parameters in these models
that are relevant to us are $\kappa_{\rm abs}$, the 
absorption coefficient of the unknown stratospheric absorber,
and $\rm P_n$, the heat redistribution parameter, which describes
the amount of stellar flux transported from the day side of the planet
to its night side. $\kappa_{\rm abs}$ has units of $\rm cm^2\,g^{-1}$, 
while $\rm P_n$ is unitless and varies between 0 (no redistribution) 
and 0.5 (complete redistribution). 

In the Fortney models, essentially, only the heat redistribution efficiency, 
$f$, is a free parameter. The high altitude TiO and VO at 
equilibrium abundances serve as absorbers causing temperature
inversions when needed to explain the data. The heat redistribution 
efficiency varies between $f = 0.25$ (indicating that the flux is 
evenly distributed over the whole planet) and $f = 0.67$ (corresponding to no 
flux redistribution at all, even within day-side regions of different 
temperature). The value $f = 0.5$ signifies that flux is evenly redistributed
on the day-side of the planet, but no heat leaks to the night side.
The atmospheres in the Fortney models have solar compositions, 
except the $f = 0.6$ HAT-P-12b model 
(shown in red in Figure~\ref{fig:models}), which has a 30$\times$ solar metallicity. 
Neither the Fortney nor Burrows models account for the presence of clouds or
disequilibrium chemical processes.

``Fitting'' these models in the mathematical sense is not practical. 
In the Fortney models one can vary $f$ and the presence or absence of 
TiO and VO, while the Burrows models have 
$\kappa_{\rm abs}$ and $\rm P_n$ as free parameters, 
but neither set of models is intended to include an algorithm to adjust these 
parameters according to the data as part of the model calculation. The best 
approach available with the current state of the model codes is 
to compute a range of models for a given planet covering the parameter
space, and then to select manually the one that accounts best 
for the data. For HAT-P-3b, we visually examine Burrows models with $\kappa_{\rm abs}$
between $0$ and $0.1\,cm^2\,g^{-1}$, and $\rm P_n$ between 
$0.1$ and $0.3$; for HAT-P-4b we look at $\kappa_{\rm abs}$ between $0$ and 
$0.2\,cm^2\,g^{-1}$, and $\rm P_n$ between $0.1$ and $0.3$; 
and for HAT-P-12b we experiment with $\kappa_{\rm abs}$ between $0$ and
$0.1\,cm^2\,g^{-1}$, keeping $\rm P_n$ at a moderate value of $0.3$. 
We also examine Fortney models with and without TiO and VO in the 
upper layers of the atmosphere, with $f$ between 0.25 and 0.6. 
Below, we discuss models that match the data well, 
and our conclusions for the atmosphere of each planet.

\subsection{HAT-P-3b and HAT-P-4b}

HAT-P-3 displays a slightly enhanced level of chromospheric activity, 
with calcium H\&K activity index $\rm log(R^\prime_{HK}) = -4.904$ \citep{knu10}. 
This could explain some of the red noise evident in the light curves
and PBMC runs for this planet (Figures~\ref{fig:bin} and \ref{fig:edhist}). 

For HAT-P-3b (top panels in Figure~\ref{fig:models}), 
we adopt the Burrows model with $\kappa_{\rm abs} = 0.1\,cm^2\,g^{-1}$ and 
$\rm P_n = 0.1$, indicating an atmosphere with a temperature 
inversion and modest heat redistribution. 
We find that the atmosphere of HAT-P-3b is matched by a 
Fortney model with $f = 0.6$. This planet, like
HAT-P-12b, is cool enough that the presence of TiO and VO has 
practically no effect on the shape of the spectra, and 
the Fortney models cannot be used to distinguish between inverted
and non-inverted temperature profiles. HAT-P-3b's eclipse depths are matched
equally well by the almost identical inverted and non-inverted spectral models. 

Both sets of models, however, agree on low redistribution efficiency, in 
apparent contradiction of the hypothesis of \citet{per12} that
hot Jupiters with irradiation temperature, $\rm T_{irr}\lesssim2000$\,K
have efficient flux redistribution. In their convention, 
$\rm T_{irr} = T_{eff}(R_\star/a)^{1/2}$, where $\rm T_{eff}$ is the 
effective temperature of the star, $\rm R_\star$ is the radius of 
the star and $\rm a$ is the semimajor axis. For HAT-P-3b, 
$\rm T_{irr}=1600$\,K. Our result does not contradict (but does not support either)
the hypothesis by \citet{cow11}, who suggest that a wide range of redistribution 
efficiencies are possible for planets with $\rm T_{\epsilon=0}\lesssim2400$\,K, 
since $\rm T_{\epsilon=0} = 1300$\,K for HAT-P-3b.

HAT-P-4 is chromospherically quiet, with an activity index 
of $\rm log(R^\prime_{HK}) = -5.082$ \citep{knu10}, and we find no
significant perturbations to the light curves other than 
the eclipses. This planet is hotter than HAT-P-3b and HAT-P-12b
and therefore has deeper eclipses, despite the fact that its
host star is more luminous than the host stars of the other two. 

The HAT-P-4b models are presented in the middle panels 
in Figure~\ref{fig:models}. We find that a Burrows model
that describes the data well has $\kappa_{\rm abs} = 0.2\,cm^2\,g^{-1}$ 
and $\rm P_n = 0.1$, corresponding to an inverted atmosphere
with inefficient heat redistribution. The Fortney models 
with and without TiO and VO absorption in the upper 
atmosphere and $f = 0.5$ also seem to be close to the observations.
As in the HAT-P-3b case, the models appear to be ambiguous
about any temperature inversions, but agree that the planet's
atmosphere has moderate to low efficiency in redistributing 
heat to the night side. 

For HAT-P-4b, $\rm T_{irr}=2400$\,K. Therefore, the inefficient flux 
redistribution we observe is consistent with the idea that planets with 
$\rm T_{irr}$ above 2200-2400\,K should have little flux transfer to 
their night sides \citep{per12,cow11}. 

\citet{knu10} hypothesize that while chromospherically
active stars have planets with non-inverted atmospheres, planets around 
quiet stars have temperature inversions. The difference between the
activity indices of the host stars HAT-P-3 and HAT-P-4 is minimal 
($\rm \Delta log(R^\prime_{HK})=0.1$), and both fall in the 
tentative border region where hot Jupiters can have inverted or non-inverted atmospheres
and so predictions for the presence or absence of an inversion are
difficult. Due to this and the ambiguity of the models of the planetary atmospheres, 
we cannot make claims that support or contradict this idea based on our 
data on these two planets. 

\subsection{HAT-P-12b}
In principle, it is possible that we have failed to detect the HAT-P-12b
eclipses due to a relatively large orbital eccentricity. The discovery 
paper by \citet{har09} fixes the eccentricity at 0, which is what we have
assumed in our determination of the upper limit of the eclipse depth. However, 
their initial fit results in a best value of $e\,\cos\omega = 0.052\pm0.025$, 
which is insignificant. 
We estimate that we would have detected eclipses centered between phases
of 0.45 and 0.55 (the range of our data in units of orbital phase), 
corresponding to $|e\,\cos\omega| < 0.08$. This is only about 
1$\sigma$ from the insignificant value by \citet{har09}.
Using the Exoplanet Orbit Database \citep[http://exoplanets.org,][]{wri11}, 
we find that out of the 188 known transiting planets with periods 
less than 10\,days, only 23 ($\sim12\%$) have orbital eccentricity over 0.08.
Thus, we conclude that it is possible that we have missed the eclipse due to 
eccentricity larger than $\sim0.08$, but that weak eclipses are a more likely 
explanation. 

Despite the formal non-detection, the HAT-P-12b 4.5\,$\mu$m 
light curve in Figure~\ref{fig:bin} appears to the eye to contain 
an eclipse. Allowing for a quadratic out-of-eclipse
variation (due to, e.g., stellar spots or phase-variation),
then we find a non-zero eclipse depth of $\sim0.049\%\pm0.021\%$ 
(a $\sim2.5\sigma$ result; still not a detection).
This result is well within one sigma of the best value in
Figure~\ref{fig:edhist}, 
but is in better agreement with the non-inverted Burrows model. However, 
none of the other light curves requires anything
but a flat out-of-eclipse baseline. Using a
quadratic curve to allow for phase variation will naturally cause any
eclipse in the data to look deeper, since the assumed maximum 
of the quadratic curve is near mid-eclipse, above what is 
expected for a flat baseline. On the other hand, the appearance of
an eclipse and a phase curve variation is likely to be caused by 
instrumental red noise, given the lack of similar signatures in
the other two 4.5\,$\mu$m curves. HAT-P-3b and HAT-P-4b are hotter than 
HAT-P-12b and have deeper eclipses, so any 
phase curve variability should be higher for them. In addition, 
the HAT-P-12 host star has a Ca~II H\&K activity index, $\rm log(R^\prime_{HK})$,
of about $-5.1$ \citep{knu10}, so stellar spots should be a less important
factor than in the light curve than they are, e.g., in the Sun 
\citep[$\rm log(R^\prime_{HK}) = -4.9$,][]{noy1984}. Hence, we conclude
that a quadratic out-of-eclipse variation does not improve our results. 

We are only able to place upper limits on the secondary eclipses of HAT-P-12b,
(assuming $|e\,\cos\omega| = 0$),
however, we can still compare these results to atmospheric models 
(lower panels in Figure~\ref{fig:models}). Both
the inverted Burrows ($\rm 0.1\,cm^2\,g^{-1}$, 
$\rm P_n = 0.3$) and the Fortney ($f = 0.5$ and $0.6$, with
30 times solar metallicity for the latter, with the presence
or absence of TiO irrelevant to the shape of the spectrum, due
to the low temperature of the planet) models are poor fits to 
the data. The non-inverted Burrows model ($\kappa_{\rm abs} = 0$, 
$\rm P_n = 0.3$) appears close to the $3\sigma$ eclipse depth 
upper limits, consistent with a lack of temperature inversion. 

The non-detections of thermal emission at both {\it Spitzer} channels, 
may suggest that heat is relatively efficiently redistributed to 
the night side of HAT-P-12b and/or that the albedo is high.
However, it is difficult to claim this with any certainty, given 
the poor match that the models provide for the HAT-P-12b measurements. 
For this planet, $\rm T_{irr}=1350$\,K, but its surface gravity is 
$\sim5.7$\,m\,s$^{-2}$, significantly less than 
the value of 10\,m\,s$^{-2}$, assumed for ``typical'' hot Jupiters by 
\citet{per12}. For these two reasons, we do not consider the possible high heat redistribution 
efficiency of HAT-P-12b to be evidence in favor of the \citet{per12} hypothesis. 

The non-detection of thermal flux from HAT-P-12b is intriguing
since neither the Burrows nor the Fortney models can account well for this.
Searching the literature, we find two other cool low-mass planets
observed with {\it Spitzer} at 3.6 and 4.5\,$\mu$m -- GJ~436b \citep{but04,gil07b,ste10}
and WASP-29b \citep{har12}. 
GJ~436b, WASP-29b and HAT-P-12b are the three coolest planets observed during
secondary eclipse with {\it Spitzer}. They have equilibrium temperatures of 
$\sim600$\,K, $\sim980$ and $\sim950$\,K, respectively, assuming complete 
redistribution of the stellar flux to the night side and zero albedo. Assuming 
no flux redistribution and zero albedo, the average equilibrium temperatures of their day 
sides are $\sim700$\,K, $\sim1110$\,K and $\sim1120$\,K, respectively. For these 
calculations, we assume that the planets are always at a distance of 1 
semi-major axis away from their host stars, which is relevant for GJ~436b, 
since it has non-zero eccentricity, $e=0.16$. GJ~436b orbits an M2.5~V star 
with effective temperature, $\rm T_{\rm eff} = 3350\pm300$\,K \citep{man07}, and 
$\rm [M/H]=-0.32\pm0.12$ \citep{bea06}. The HAT-P-12b and WASP-29b host 
stars are larger -- K4 dwarfs with  $\rm T_{\rm eff} = 4650\pm60$\,K, 
and $4800\pm150$\,K, respectively. The metallicity of HAT-P-12 is almost identical to 
that of GJ~436: [Fe/H]$ = -0.29 \pm 0.05$, but for WASP-29, [Fe/H]$ = 0.11\pm0.014$
\citep{har09, hel10}. WASP-29b, like HAT-P-12b has a mass similar to Saturn's, 
while GJ~436b is a hot Neptune. 

WASP-29b and GJ~436b exhibit a measurable eclipse depth at 
3.6\,$\mu$m, while none is detected at 4.5\,$\mu$m. This, in 
combination with {\it Spitzer} secondary eclipse 
measurements for GJ~436b at 5.8, 8.0, 16 and 24\,$\mu$m, prompts \citet{ste10}
to suggest that the planet has a non-inverted atmosphere with 
large concentrations of CO at the expense of $\rm CH_4$. 
They argue that the small eclipse depth at 
4.5\,$\mu$m is a result of a strong absorption feature of CO, while
the strong 3.6\,$\mu$m eclipse is caused by lack of $\rm CH_4$ 
absorption. $\rm CH_4$ is expected to begin to dominate as a carbon 
bearing molecule below temperatures of about 1100\,K 
(for pressures of 1\,bar and solar metallicity) \citep{lod02, for08}, 
and so the suggested abundance of CO would require 
thermochemical disequilibrium in the GJ~436b's atmosphere.
\citet{har12} suggest that a similar explanation is possible for WASP-29b. 

Despite their similar host stars, irradiation levels, planetary masses and
planetary equilibrium temperatures, HAT-P-12b exhibits a completely 
different behavior from WASP-29b, since it produces prominent 
eclipses at neither 3.6, nor 4.5\,$\mu$m. The reasons for this disparity are unclear
and additional atmospheric modeling and extensive observations are needed to 
investigate the atmospheres of planets with temperatures $\lesssim1000$\,K, 
which appear to be very dissimilar to those of traditional hot Jupiters. 

\subsection{Orbital Eccentricity Constraints}

The measured time of secondary eclipse can be used to constrain 
the orbital eccentricity, $e$, as part of the quantity 
$|e\,\cos\omega|$, where $\omega$ is the argument of 
periastron. For a detailed discussion, see, e.g., \citet{cha05}. For HAT-P-3b
and HAT-P-4b, we average the timed eclipse central phases, weighing them
by the inverse of their variance and derive central phases of 
$0.50292\pm0.00076$ (HAT-P-3b) and $0.49951\pm0.00070$ (HAT-P-4b). 
The individual timings for HAT-P-4b agree within 0.1\,$\sigma$, but
the ones for HAT-P-3b disagree at the 2.8\,$\sigma$ level. We suggest 
that this discrepancy may be due to residual red noise, potentially due to
stellar activity artifacts in the light curves for this planet that has 
not been taken into account by our data analysis procedures. 

Even for a circular orbit, or for an argument of periastron of 0/180$^{\circ}$, 
the central phase of the eclipse is not expected to be exactly 0.5. 
The light travel time delay is 19\,s, 22\,s and 19\,s, corresponding
to to expected central phases of 0.50008, 0.50008 and 0.50007 for HAT-P-3b, 
HAT-P-4b and HAT-P-12b, respectively. Another source of apparent 
delay could be an off-centered hottest point on the planet's face,
closer to the trailing side of the planet than to its leading limb  
\citep[e.g., ][]{cha05, coo05, knu07, agol10}. Orbital perturbations
due to previously undetected planets can cause the secondary eclipse to occur 
earlier or later than expected. Our timing precision is not sufficient to 
detect any of these effects.

Taking the light travel time delay into account, and following the 
discussion in \citet{cha05}, we find that for HAT-P-3b, $|e\,\cos\omega| < 0.0081$ 
within 3$\sigma$. Similarly, within 3$\sigma$, $|e\,\cos\omega| < 0.0042$
for HAT-P-4b. These values are consistent with the value of 
$e = 0$ adopted by the discovery studies \citep{tor07, kov07} for 
all three planets. Since we do not detect the eclipses of HAT-P-12b, we cannot 
constrain its eccentricity.

\section{Conclusion}
We measure secondary eclipse depths in the 3.6 and 4.5\,$\mu$m bands
of {\it Warm Spitzer} for three exoplanets, HAT-P-3b, HAT-P-4b and
HAT-P-12b. We find that HAT-P-3b and HAT-P-4b have inefficient heat transfer from 
their day to night sides, but the models we compare to the data
are ambiguous with respect to possible temperature inversions in their 
atmospheres. We detect the eclipses of HAT-P-12b neither at 4.5\,$\mu$m, 
nor at 3.6\,$\mu$m. This result is in contrast with 
{\it Spitzer} eclipse measurements of hot Neptune GJ~436b by \citet{ste10} 
and hot Saturn WASP-29b by \citep{har12}, two planets in the same temperature 
regime as HAT-P-12b. This is a confirmation that current models need 
further development to explain observations of planets with 
effective temperatures below $~1200$\,K. Additional infra-red secondary eclipse 
observations of ``warm Jupiters'' are urgently needed to constrain these models. 
Even non-detections, as in this study, combined with accurate estimates
of the $|e\,\cos\omega|$ quantity would be invaluable in assessing the plausibility 
of atmospheric models. Currently, the {\it Spitzer Space Telescope} remains the best 
available observatory to perform these studies. 

In addition, we find that the eccentricities of HAT-P-3b and HAT-P-4b  
are consistent with zero. This is in agreement with previously published 
radial velocity measurements. We compile transit timing measurements 
from the literature and improve the ephemeris of HAT-P-12b.
We see no obvious transit timing variations that could indicate additional objects 
in the HAT-P-12b system.   

Future eclipse depth observations will benefit from the experience 
that the community has gained from the study of exoplanets with the {\it Warm Spitzer} mission.
The pointing oscillation observed for long time series observations with IRAC has 
been a problem because it causes apparent brightness changes of the 
target stars due to intrapixel response variations (the pixel-phase effect).
This effect has been mitigated, for observations after October 17, 2010, by adjusting the operation of a heater
designed to keep a battery within its operating temperature range. Both
the amplitude and the period of the pointing wobble have been reduced, resulting 
in smaller brightness changes over shorter periods (from $\sim60$\,min to 
$\sim40$\,min), making them more distinguishable from astrophysical 
phenomena like exoplanet transits and eclipses. As a result, the removal 
of the pointing oscillation effect can be done more efficiently and
precisely\footnote{\scriptsize http://ssc.spitzer.caltech.edu/warmmission/news/21oct2010memo.pdf}.
In December 2011, improved pointing using the Pointing Calibration and Reference Sensor 
(PCRS) was added to the warm-mission IRAC time series photometric observations. This
option was available and frequently used during the cryogenic mission and, when it is possible 
to utilize it, can be expected to help reduce the pixel-phase effect for staring observations of 
point sources longer than $\sim12$\,hours\footnote{\scriptsize 
http://irsa.ipac.caltech.edu/data/SPITZER/docs/irac/pcrs\_obs.shtml}.
These improvements increase the value of the observatory as one of the best 
currently available tools to study exoplanet atmospheres.

\acknowledgements
We thank Jonathan Fraine for helpful discussions on ``prayer bead''
uncertainty estimation. This work is based on observations made with the 
{\it Spitzer Space Telescope}, which is operated by the Jet Propulsion 
Laboratory, California Institute of Technology under a contract 
with NASA. Support for this work was provided by NASA through an 
award issued by JPL/Caltech.

\newpage
\clearpage
\center{
\begin{deluxetable}{llll}
\tabletypesize{\scriptsize}
\tablewidth{0pt}
\tablecaption{Adopted Stellar and Planetary Parameters}
\tablehead{
\colhead{} &
\colhead{HAT-P-3b\tablenotemark{d}} &
\colhead{HAT-P-4b\tablenotemark{e}} &
\colhead{HAT-P-12b\tablenotemark{g}}} 
\startdata
M$_\star$ (M$_{\Sun}$) & $0.917 \pm 0.030$ & $1.271^{+0.120}_{-0.070}$           & $0.733 \pm 0.018$ \\
R$_\star$ (R$_{\Sun}$) & $0.799 \pm 0.039$ & $1.600^{+0.117}_{-0.042}$           & $0.701^{+0.017}_{-0.012}$ \\
$\rm K_s$ (mag)\tablenotemark{a}& $9.448 \pm 0.025$ & $9.770 \pm 0.020$     & $10.108 \pm 0.016$\\
T$_{\rm eff}$ (K)      & $5185 \pm 80$     & $5990$\tablenotemark{f}          & $4650$\tablenotemark{f} \\
b$_{\rm impact}$       & $0.530 \pm 0.075$ & $0.084^{+0.014}_{-0.026}$            & $0.211 \pm 0.012$ \\
M$_{\rm p}$ (M$_{\rm J}$)    & $0.591 \pm 0.018$ & $0.680^{+0.038}_{-0.025}$       & $0.211^{+0.066}_{-0.078}$ \\
R$_{\rm p}$ (R$_{\rm J}$)    & $0.827 \pm 0.055$ & $1.337^{+0.079}_{-0.036}$       & $0.959^{+0.029}_{-0.021}$ \\
P (days)\tablenotemark{b} & $2.8997382 \pm 0.0000009$ & $3.0565254 \pm 0.0000012$ & $3.21305929 \pm 0.00000034$ \\
a$_{\rm p}$ (AU)\tablenotemark{c} & $0.03866 \pm 0.00042$ & $0.04465 \pm 0.00111$    & $0.03842 \pm 0.00031$
\enddata
\tablenotetext{a}{Two Micron All Sky Survey (2MASS) $\rm K_s$ magnitude of the star.}
\tablenotetext{b}{The HAT-P-3b and HAT-P-4b orbital periods are taken \citet{sad12}; 
the HAT-P-12b period is derived from our updated ephemerides shown in Table~\ref{tab:eph} and Figure~\ref{fig:eph}.}
\tablenotetext{c}{Calculated from the stellar masses and orbital periods assumed in this table.}
\tablenotetext{d}{Values from \citet{chan11}, except for the magnitude, $\rm K_s$, the period, P, and the semimajor axis.}
\tablenotetext{e}{Values from \citet{sou11}, except for the magnitude, $\rm K_s$, the effective temperature, T$_{\rm eff}$, the period, P, and the semimajor axis.}
\tablenotetext{f}{\citet{knu10}.}
\tablenotetext{g}{Values from \citet{har09}, except for the magnitude, $\rm K_s$, the effective temperature, T$_{\rm eff}$, the period, P, and the semimajor axis.}
\label{tab:prop}
\end{deluxetable}
}

\center{
\begin{deluxetable}{lllllll}
\tabletypesize{\scriptsize}
\tablewidth{0pt}
\tablecaption{Photometry Parameters}
\tablehead{
\colhead{} &
\multicolumn{2}{c}{HAT-P-3b} &
\multicolumn{2}{c}{HAT-P-4b} &
\multicolumn{2}{c}{HAT-P-12b} \\
\cline{2-7}
\colhead{} &
\colhead{$3.6\,\mu$m} &  
\colhead{$4.5\,\mu$m} &  
\colhead{$3.6\,\mu$m} &   
\colhead{$4.5\,\mu$m} &   
\colhead{$3.6\,\mu$m} &  
\colhead{$4.5\,\mu$m}}
\startdata
$\rm r_{fix}$ (pixels)\tablenotemark{a} & 2.5  & 3.0  & 4.0  & 3.0  & 5.0  & 2.5 \\
\hline
$\rm R_I$ (pixels)\tablenotemark{b}    & 3.5  & 6.5  & 4.0  & 6.5  & 6.0  & 6.5  \\
b \tablenotemark{c}                    & 0.40 & 0.30 & 0.85 & 0.50 & 0.95 & 0.50 \\
c (pixels) \tablenotemark{c}           & 1.30 & 1.60 & 0.35 & 1.05 & 1.05 & 1.10 \\
median $\rm r_{var}$ \tablenotemark{d}  & 2.21 & 2.33 & 2.24 & 2.55 & 3.10 & 2.46 \\
maximum $\rm r_{var}$\tablenotemark{d}  & 2.33 & 2.39 & 2.47 & 2.64 & 3.24 & 2.58 \\
minimum $\rm r_{var}$\tablenotemark{d}  & 2.16 & 2.25 & 2.09 & 2.41 & 2.96 & 2.37
\enddata
\tablenotetext{a}{Radius adopted for fixed aperture photometry.}
\tablenotetext{b}{Radius adopted for calculating the stellar intensity for Equation~\ref{eqn:beta}.}
\tablenotetext{c}{Variable aperture photometry parameters adopted, see Equation~\ref{eqn:beta}.}
\tablenotetext{d}{Median, maximum and minimum variable aperture radius.}
\label{tab:phot}
\end{deluxetable}}

\center{
\begin{deluxetable}{lllllll}
\tabletypesize{\scriptsize}
\tablewidth{0pt}
\tablecaption{Observation Details}
\tablehead{
\colhead{} &
\multicolumn{2}{c}{HAT-P-3b} &
\multicolumn{2}{c}{HAT-P-4b} &
\multicolumn{2}{c}{HAT-P-12b} \\
\cline{2-7}
\colhead{} &
\colhead{$3.6\,\mu$m} &  
\colhead{$4.5\,\mu$m} &  
\colhead{$3.6\,\mu$m} &   
\colhead{$4.5\,\mu$m} &   
\colhead{$3.6\,\mu$m} &  
\colhead{$4.5\,\mu$m}}
\startdata
Observation start (UTC)  & 17-03-2010,02:45  & 20-03-2010,00:04 & 12-04-2010,02:21 & 02-09-2010,18:49 & 16-03-2010,13:58
& 26-03-2010,05:09\\
Observation end   (UTC)  & 17-03-2010,10:28  & 20-03-2010,07:46 & 12-04-2010,09:59 & 03-09-2010,02:27 & 16-03-2010,21:35 
& 26-03-2010,12:47 \\
Orbital phase coverage   & 0.433 -- 0.544  & 0.428 -- 0.539 & 0.439 -- 0.543 & 0.448 -- 0.552 & 0.445 -- 0.544 
& 0.443 - 0.542 \\
Image count              & 13,760           & 13,760         & 3,871            & 3,871            & 2,097
 & 2,097
\enddata
\label{tab:dat}
\end{deluxetable}}

\clearpage
\renewcommand{\arraystretch}{1}
\begin{deluxetable*}{lcllc}
\tabletypesize{\scriptsize}
\tablewidth{0pt}
\tablecaption{HAT-P-12b Ephemerides}
\tablehead{
\colhead{N} &\colhead{Observation Date} &\colhead{Primary Transit ($\rm BJD_{TT}$)} & \colhead{Uncertainty} & \colhead{Notes}}
\startdata
0   & 2007 Mar 28 & 2454187.85655 & 0.00020 & \citet{har09}  \\ 
9   & 2007 Apr 26 & 2454216.77265 & 0.00014 & \citet{har09}  \\ 
212 & 2009 Feb 06 & 2454869.02397 & 0.00017 & \citet{har09}  \\ 
221 & 2009 Mar 07 & 2454897.94225 & 0.00024 & \citet{har09}  \\ 
238 & 2009 May 01 & 2454952.56398 & 0.00080 & ETD \\
242 & 2009 May 13 & 2454965.41639 & 0.00046 & ETD \\ 
242 & 2009 May 13 & 2454965.41748 & 0.00090 & ETD \\
248 & 2009 Jun 02 & 2454984.69368 & 0.00060 & ETD \\
350 & 2010 Apr 25 & 2455312.42673 & 0.00032 & ETD \\
361 & 2010 May 31 & 2455347.76929 & 0.00021 & \citet{sad12} \\
449 & 2011 Mar 10 & 2455630.51896 & 0.00049 & ETD \\
454 & 2011 Mar 26 & 2455646.58477 & 0.00059 & ETD \\
454 & 2011 Mar 26 & 2455646.58486 & 0.00040 & ETD \\
455 & 2011 Mar 29 & 2455649.79769 & 0.00020 & \citet{lee12} \\
458 & 2011 Apr 07 & 2455659.43563 & 0.00038 & ETD \\
460 & 2011 Apr 14 & 2455665.86234 & 0.00031 & \citet{lee12} \\
463 & 2011 Apr 23 & 2455675.49947 & 0.00064 & ETD \\
464 & 2011 Apr 27 & 2455678.71445 & 0.00045 & \citet{sad12} \\ 
469 & 2011 May 13 & 2455694.78089 & 0.00024 & \citet{lee12} \\
472 & 2011 May 22 & 2455704.42185 & 0.00038 & ETD \\
\sidehead{These 20 transits yield:}
\sidehead{$\rm P = 3.21305929 \pm 0.00000034$\,days}
\sidehead{$\rm T_0 = 2454187.85559 \pm 0.00011$ in $\rm BJD_{TT}$}
\sidehead{{\bf Note:} ETD transits were taken from the Exoplanet Transit Database, 
http://var2.astro.cz/ETD/ and compiled by \citet{lee12}.}
\sidehead{For observations after 31 Dec 2008 and before 30 June 2012, 
$\rm TT \approx UTC + $66.184$\,s$, while for the two 2007 transits} 
\sidehead{$\rm TT \approx UTC + $65.184$\,s$.}
\enddata
\label{tab:eph}
\end{deluxetable*}
\renewcommand{\arraystretch}{1.0}

\renewcommand{\arraystretch}{2}
\begin{deluxetable}{lccccc}
\tabletypesize{\scriptsize}
\tablewidth{0pt}
\tablecaption{Secondary Eclipse Results}
\tablehead{
\colhead{}&
\colhead{Eclipse Depth (\%)} &
\colhead{Brightness Temperature (K)\tablenotemark{a}}&
\colhead{Eclipse Central Phase} &
\colhead{${\rm BJD_{\rm TT}\tablenotemark{b}-2\,450\,000}$} &
\colhead{$\rm O-C$\tablenotemark{c} (min)}
}
\startdata
HAT-P-3b, $3.6\,\mu$m & $0.112^{+0.015}_{-0.030}$ & $1575^{+75}_{-162}$ & $0.50515^{+0.00092}_{-0.00110}$ & $5272.82936^{+0.00264}_{-0.00316}$ & $21.2^{+3.8}_{-4.6}$ \\
HAT-P-3b, $4.5\,\mu$m & $0.094^{+0.016}_{-0.009}$ & $1268^{+77}_{-45}$ &  $0.50084^{+0.00106}_{-0.00071}$ & $5275.71660^{+0.00304}_{-0.00203}$ & $3.2^{+4.4}_{-3.0}$\\
HAT-P-4b, $3.6\,\mu$m & $0.142^{+0.014}_{-0.016}$ & $2194^{+98}_{-116}$ & $0.49945^{+0.00091}_{-0.00081}$ & $5298.78653^{+0.00275}_{-0.00245}$ & $-2.8^{+4.0}_{-3.5}$\\
HAT-P-4b, $4.5\,\mu$m & $0.122^{+0.012}_{-0.014}$ & $1819^{+83}_{-100}$ & $0.49960^{+0.00110}_{-0.00102}$ & $5442.44368^{+0.00333}_{-0.00309}$ & $-2.1^{+4.8}_{-4.5}$\\
HAT-P-12b, $3.6\,\mu$m & $< 0.042$ & $< 970$ & \nodata & \nodata & \nodata \\
HAT-P-12b, $4.5\,\mu$m & $< 0.085$ & $< 980$ & \nodata & \nodata & \nodata
\enddata
\tablenotetext{a}{The uncertainty of the brightness temperature included here only takes into account
the uncertainty of the eclipse depths, but not the uncertainties in the stellar properties and the 
planetary radius.}
\tablenotetext{b}{Time of secondary eclipse central phase, 
in Barycentric Julian Date (BJD) based on Terrestrial Time  (TT). For these observations, 
$\rm TT \approx UTC + 66.184$\,s, where UTC is the Universal Coordinated Time.}
\tablenotetext{c}{The measured offset from the expected central phase of 0.50008 
(HAT-P-3b and HAT-P-4b), with an adjustment for light travel time, (see Section~\ref{sec:eph}), in minutes.}
\label{tab:ed}
\end{deluxetable}
\renewcommand{\arraystretch}{1.0}


\begin{thebibliography}{}
\bibitem[Agol et al.(2010)]{agol10}
Agol, E., Cowan, N. B., Knutson, H. A., Deming, D., Steffen, J. H.,
Henry, G. W., \& Charbonneau, D. 2010, \apj, 721, 1861

\bibitem[Anderson et al.(2011)]{and11} 
Anderson, D.~R., Smith, A.~M.~S., Lanotte, A.~A., et al.\ 2011, \mnras, 416, 2108 

\bibitem[Ballard et al.(2010a)]{bal10} 
Ballard, S., Charbonneau, D., Deming, D., et al.\ 2010, \pasp, 122, 1341 

\bibitem[Bean et al.(2006)]{bea06} 
Bean, J.~L., Benedict, G.~F., \& Endl, M.\ 2006, \apjl, 653, L65 

\bibitem[Beerer et al.(2011)]{bee11}
Beerer, I.~M., et al. 2011, \apj, 727, 23

\bibitem[Burrows \& Sharp(1999)]{bur99} 
Burrows, A., \& Sharp, C.~M.\ 1999, \apj, 512, 843 

\bibitem[Burrows et al.(2007)]{bur07}
Burrows, A., Hubeny, I., Budaj, J., Knutson, H.~A., 
\& Charbonneau, D. 2007, \apjl, 668, L171 

\bibitem[Burrows et al.(2008)]{bur08}
Burrows, A., Budaj, J., \& Hubeny, I. 2008, \apj, 678, 1436 

\bibitem[Butler et al.(2004)]{but04} 
Butler, R.~P., Vogt, S.~S., Marcy, G.~W., et al.\ 2004, \apj, 617, 580 

\bibitem[Campo et al.(2011)]{cam11} 
Campo, C.~J., et al. 2011, \apj, 727, 125 

\bibitem[Chan et al.(2011)]{chan11} 
Chan, T., Ingemyr, M., Winn, J.~N., et al.\ 2011, \aj, 141, 179 

\bibitem[Charbonneau et al.(2005)]{cha05}
Charbonneau, D., et al. 2005, ApJ, 626, 523

\bibitem[Charbonneau et al.(2008)]{cha08} 
Charbonneau, D., Knutson, H.~A., Barman, T., Allen, L.~E., 
Mayor, M., Megeath, S. T., Queloz, D., \& Udry, S. 2008, \apj, 686, 1341

\bibitem[Christiansen et al.(2010)]{chr10} 
Christiansen, J.~L., et al. 2010, \apj, 710, 97 

\bibitem[Cochran et al.(2011)]{coc11} 
Cochran, W.~D., et al. 2011, \apjs, 197, 7 

\bibitem[Cooper \& Showman(2005)]{coo05}
Cooper, C. S. \& Showman, A. P. 2005, \apjl, 629, L45

\bibitem[Correia \& Laskar(2010)]{cor10}
{Correia}, A.~C.~M. \& {Laskar}, J. 2010, in Exoplanets, ed. S.~Seager,
(Tucson: Univ. Arizona Press), 239

\bibitem[Cowan \& Agol(2011)]{cow11} 
{Cowan}, N.~B., \& {Agol}, E. 2011, \apj, 729, 54

\bibitem[Cowan et al.(2012)]{cow12} 
Cowan, N.~B., Machalek, P., Croll, B., et al.\ 2012, \apj, 747, 82 

\bibitem[Deming et al.(2005)]{dem05}
{Deming}, D., {Seager}, S., {Richardson}, \& 
L.~J., {Harrington}, J. 2005, Nature, 434, 740

\bibitem[Deming et al.(2011)]{dem11} 
{Deming}, D., et al. 2011, \apj, 726, 95

\bibitem[Demory et al.(2011)]{dmr11} 
Demory, B.-O., et al. 2011, \aap, 533, A114 

\bibitem[D{\'e}sert et al.(2011)]{des11} 
D{\'e}sert, J.-M., Charbonneau, D., Fortney, J.~J., et al.\ 2011, \apjs, 197, 11 

\bibitem[Eastman et al.(2010)]{eas10} 
{Eastman}, J., {Siverd}, R. \& {Gaudi}, B.~S. 2010, \pasp, 122, 935

\bibitem[Fazio et al.(2004)]{faz04} 
Fazio, G. G., et al. 2004, ApJS, 154, 10

\bibitem[Ford(2005)]{ford05} 
Ford, E.~B.\ 2005, \aj, 129, 1706 

\bibitem[Ford(2006)]{ford06} 
Ford, E.~B.\ 2006, \apj, 642, 505 

\bibitem[Fortney et al.(2005)]{for05}
Fortney, J. J., Marley, M. S., Lodders, K., Saumon, D., \&
Freedman, R. S. 2005, ApJ, 627, L69

\bibitem[Fortney et al.(2006a)]{for06a}
Fortney, J. J., Saumon, D., Marley, M. S., Lodders, K., \&
Freedman, R. S. 2006a, ApJ, 642, 495

\bibitem[Fortney et al.(2006b)]{for06b}
Fortney, J. J., et al. 2006b, ApJ, 652, 746

\bibitem[Fortney et al.(2008)]{for08}
Fortney, J. J., Lodders, K., Marley, M. S., \& Freedman, R. S. 2008, ApJ, 678, 1419

\bibitem[Fressin et al.(2010)]{fre10}
Fressin, F., et al. 2010, ApJ, 711, 374

\bibitem[Gillon et al.(2007a)]{gil07a}
Gillon, M., et al. 2007, \aap, 471, L51

\bibitem[Gillon et al.(2007b)]{gil07b} 
Gillon, M., Pont, F., Demory, B.-O., et al.\ 2007, \aap, 472, L13 

\bibitem[Grillmair et al.(2008)]{gri08}
Grillmair, C.~J., et al. 2008, \nat, 456, 767

\bibitem[Hardin et al.(2012)]{har12} Hardin, M., Harrington, 
J., Stevenson, K., et al.\ 2012, AAS/Division for Planetary Sciences 
Meeting Abstracts, 44, \#200.09 

\bibitem[Harrington et al.(2007)]{har07} 
Harrington, J., Luszcz, S., Seager, S., Deming, D., 
\& Richardson, L.~J.\ 2007, \nat, 447, 691 

\bibitem[Hartman et al.(2009)]{har09} 
Hartman, J.~D., Bakos, G.~{\'A}., Torres, G., et al.\ 2009, \apj, 706, 785 

\bibitem[Hellier et al.(2010)]{hel10} 
Hellier, C., Anderson, D.~R., Collier Cameron, A., et al.\ 2010, \apjl, 723, L60 

\bibitem[Hubeny et al.(2003)]{hub03}
Hubeny, I., Burrows, A., \& Sudarsky, D. 2003, ApJ, 594, 1011

\bibitem[Knutson et al.(2007)]{knu07}
Knutson, H.~A, et al. 2007, \nat, 447, 183

\bibitem[Knutson et al.(2008)]{knu08} 
Knutson, H.~A., Charbonneau, D., Allen, L.~E., 
Burrows, A., \& Megeath, S.~T. 2008, \apj, 673, 526

\bibitem[Knutson et al.(2009)]{knu09}
Knutson, H.~A., Charbonneau, D., Burrows, A., 
O'Donovan, F.~T., \& Mandushev, G. 2009, \apj, 691, 866

\bibitem[Knutson et al.(2010)]{knu10} 
Knutson, H.~A., Howard, A.~W., \& Isaacson, H. 2010, \apj, 720, 1569

\bibitem[Knutson et al.(2012)]{knu12} 
Knutson, H.~A., Lewis, N., Fortney, J.~J., et al.\ 2012, \apj, 754, 22 

\bibitem[Kov{\'a}cs et al.(2007)]{kov07} 
Kov{\'a}cs, G., Bakos, G.~{\'A}., Torres, G., et al.\ 2007, \apjl, 670, L41 

\bibitem[Kurucz(1979)]{kur79}
Kurucz, R. L. 1979, ApJS, 40, 1

\bibitem[Lee et al.(2012)]{lee12} 
Lee, J.~W., Youn, J.-H., Kim, S.-L., Lee, C.-U., \& Hinse, T.~C.\ 2012, \aj, 143, 95 

\bibitem[Lewis et al.(2013)]{lew13}
Lewis, N.~K., Knutson, H.~A., Showman, A.~P., et al.\ 2013, \apj, 766, 95 

\bibitem[Lodders \& Fegley(2002)]{lod02} 
Lodders, K., \& Fegley, B.\ 2002, Icarus, 155, 393 

\bibitem[Machalek et al.(2008)]{mac08}
Machalek, P., et al. 2008, ApJ, 684, 1427

\bibitem[Machalek et al.(2009)]{mac09}
Machalek, P., McCullough, P.~R., Burrows, A., Burke, C.~J., 
Hora, J.~L., \& Johns-Krull, C.~M. 2009, \apj, 701, 514


\bibitem[Maness et al.(2007)]{man07} 
Maness, H.~L., Marcy, G.~W., Ford, E.~B., et al.\ 2007, \pasp, 119, 90 

\bibitem[Mighell(2005)]{mig05} 
Mighell, K.~J.\ 2005, \mnras, 361, 861 

\bibitem[Noyes et al.(1984)]{noy1984} 
{Noyes}, R.~W., {Hartmann}, L.~W., {Baliunas}, S.~L., 
{Duncan}, D.~K., \& {Vaughan}, A.~H. 1984, \apj, 279, 763

\bibitem[Parmentier et al.(2013)]{per13} 
Parmentier, V., Showman, A.~P., \& Lian, Y.\ 2013, \aap, submitted

\bibitem[Perna et al.(2012)]{per12} 
Perna, R., Heng, K., \& Pont, F.\ 2012, \apj, 751, 59 

\bibitem[Sada et al.(2012)]{sad12} 
Sada, P.~V., Deming, D., Jennings, D.~E., et al.\ 2012, \pasp, 124, 212 

\bibitem[Sharp \& Burrows(2007)]{sha07} 
Sharp, C.~M., \& Burrows, A.\ 2007, \apjs, 168, 140 

\bibitem[Southworth(2011)]{sou11} 
Southworth, J.\ 2011, \mnras, 417, 2166 

\bibitem[Spiegel et al.(2009)]{spi09}
Spiegel, D.~S., Silverio, K. \& Burrows, A. 2009, \apj, 699, 1487

\bibitem[Stevenson et al.(2010)]{ste10} 
Stevenson, K.~B., et al.\ 2010, \nat, 464, 1161 

\bibitem[Stevenson et al.(2012)]{ste12} 
Stevenson, K.~B., Harrington, J., Fortney, J.~J., et al.\ 2012, \apj, 754, 136 

\bibitem[Todorov et al.(2010)]{tod10}
Todorov, K.~O., Deming, D., Harrington, J., Stevenson, K. B., Bowman, W. C., 
Nymeyer, S., Fortney, J. J., \& Bakos, G. A. 2010, \apj, 708, 498

\bibitem[Todorov et al.(2012)]{tod12} Todorov, K.~O., Deming, 
D., Knutson, H.~A., et al.\ 2012, \apj, 746, 111 

\bibitem[Torres et al.(2007)]{tor07} 
Torres, G., Bakos, G.~{\'A}., Kov{\'a}cs, G., et al.\ 2007, \apjl, 666, L121 

\bibitem[Wright et al.(2011)]{wri11} 
Wright, J.~T., Fakhouri, O., Marcy, G.~W., et al.\ 2011, \pasp, 123, 412 

\bibitem[Zahnle et al.(2009)]{zah09}
Zahnle, K., Marley, M. S., Freedman, R. S., Lodders, K., \& Fortney, J. J. 2009, ApJ, 701, L20

\end{thebibliography}
\end{document}